\documentclass[showpacs,twocolumn]{revtex4}

\usepackage{amssymb}
\usepackage{amsmath}
\usepackage{array}
\usepackage{graphicx}

\def\cn{\;}

\def\adj{^\dagger}
\renewcommand\vec[1]{{\mathbf #1}}

\def\kb{k_{\mathrm{B}}}

\def\conj{^*}

\def\ph{\varphi}
\def\z{\textit{z}}
\def\mb{\mu_{\mathrm{B}}}
\def\kink{e_{\vec{k}}}

\def\Na{${}^{23}\mathrm{Na}$ }
\def\diag{\mathrm{diag}}

\def\Gr{\mathcal{G}}
\def\Gro{\mat{\Gr}{}_{(0)}}

\def\ktau{(\vec{k},\tau)}
\def\komega{(\vec{k},i\omega{_n})}
\def\Tc{T_{\mathrm{c}}}
\def\BC{B_{\mathrm{c}}}
\def\hol{\hslash\omega_{\mathrm{L}}}
\def\holt{\hslash\tilde\omega_{\mathrm{L}}}
\def\ol{\omega_{\mathrm{L}}}
\def\olt{\tilde\omega_{\mathrm{L}}}
\def\D{\Xi}
\def\WD{\Xi\conj}
\def\eH{\epsilon}
\def\cn{\mathcal{C}_n}
\def\cs{\mathcal{C}_s}
\def\Pio{\Pi_{(0)}}
\def\Pior{\Pi_{(0)}^{\text{ret}}}
\def\ps{\phi_\text{s}}

\def\xc{x_{\text{c}}}

\def\nc{n_{\text{c}}}
\def\ncp{n_{\text{c,+}}}
\def\ncm{n_{\text{c,-}}}
\def\xiB{\xi_{\text{B}}}

\newcommand\avg[1]{\left<#1\right>}
\newcommand\mat[1]{\underline{\underline{#1}}}

\allowdisplaybreaks

\begin{document}
\title{Phases of a polar spin-1 Bose gas in a magnetic
  field}
\author{Kriszti{\'a}n Kis-Szab{\'o}}
\affiliation{Department of Physics of Complex Systems, Roland E{\"o}tv{\"o}s
University, P{\'a}zm{\'a}ny P{\'e}ter s{\'e}t{\'a}ny 1/A, Budapest, H-1117}
\email{kisszabo@complex.elte.hu}
\author{P{\'e}ter Sz{\'e}pfalusy}
\affiliation{Department of Physics of Complex Systems, Roland E{\"o}tv{\"o}s
University, P{\'a}zm{\'a}ny P{\'e}ter s{\'e}t{\'a}ny 1/A, Budapest, H-1117}
\affiliation{Research Institute for Solid State Physics and Optics of the
Hungarian Academy of Sciences, Budapest, P.O.Box 49, H-1525}
\email{psz@complex.elte.hu}
\author{Gergely Szirmai}
\affiliation{Research Group for Statistical Physics of the Hungarian Academy
  of Sciences, P{\'a}zm{\'a}ny P{\'e}ter S{\'e}t{\'a}ny 1/A, Budapest, H-1117}
\email{szirmai@complex.elte.hu}
\date\today
\begin{abstract}
  The two Bose--Einstein condensed phases of a polar spin-1 gas at
  nonzero magnetizations and temperatures are investigated.
  The Hugenholtz--Pines theorem is generalized to this system.
  Crossover to a quantum phase transition is  also studied.
  Results are discussed in a mean field approximation.
\end{abstract}

\maketitle

Bose--Einstein condensed spin-1 gases have rich magnetic properties
\cite{Sea2,Stamper-Kurn2001a,GK1,KSzSz,SzkSz}. They can be classified
according to the sign of the spin-dependent part of the interaction.
In this paper this sign is assumed to be positive, i.e. the interaction is of antiferromagnetic
type.  In the absence of a magnetic field the system at zero
temperature is in the polar state \cite{Ho2,OM} and no spontaneous
magnetization is created.  An external magnetic field induces a
magnetization and as shown by Ohmi and Machida \cite{OM} in
the Bogoliubov approximation a complete spin order appears when
the magnetic field is strong enough.

The purpose of the present paper is to extend the investigations for a
homogeneous system to finite temperatures below the  Bose--Einstein condensation
(BEC). We shall denote by P2 the phase originated from the polar state when a
small magnetic field is switched on. This phase goes over to another
one denoted by P1 when the magnetic field $B$ and/or the temperature is
raised.  In the P2 phase two continuous symmetries are broken while in
phase P1 only one.  The ground state spinor has two and one
components, respectively.  Correspondingly, two gapless (Goldstone)
modes are expected to exist in P2. One of them develops a gap when
entering the phase P1.  Extended Hugenholtz--Pines theorems will be
presented whose fulfilment is a necessary condition for such a behaviour. Another
interesting aspect of this phase transition is the crossover from a
classical to a quantum phase transition when the temperature goes to
zero, which will also be investigated. Furthermore as in
\cite{Sea2,Stamper-Kurn2001a} a nonzero magnetic chemical potential
$\mu_m$ is included in the treatment. The discussion above remains valid,
if the magnetic field $B$ were replaced by $\tilde B=B+\mu_m$, which can be
regarded as an effective field determining the magnitude of the average magnetization.

For demonstration we choose a model which bears many features of
the full microscopic description. As a matter of fact, it has been found in the case of a
scalar Bose gas that the dielectric formalism can serve as a guide to
find approximations which meet conservation laws and ensure that
excitation branches associated with Goldstone modes are gapless
\cite{Griffin,FRSzG}.  Such an approach will be adopted here in a
necessarily extended form since collective motions now include
different types of spin waves besides particle density oscillations.
Further differences arise due to the presence of the effective magnetic field $\tilde B$.
Different regions are distinguished in the model on the $\tilde B-T$ plane. Properties are
discussed particularly in those regions, which might be experimentally
accessible (note that the theory worked out for a homogeneous system
can be applied for a large system as a local density approximation).

We consider a translationally invariant system of spin-1 particles in
a box with volume $V$ in a homogeneous, external magnetic field
pointing to the \z-direction.  The Hamiltonian takes the following
form:
\begin{multline}
\label{eq:ham}
  {\mathcal H}=\sum_{\genfrac{}{}{0pt}{2}{\vec{k}}{r,s}}
  \Big[(e_{\vec{k}}-\mu)\delta_{rs} -\holt\,
  (F_z)_{rs}\Big] a_r^\dagger(\vec{k})
  a_s(\vec{k})\\+\frac{1}{2V}\sum_{\genfrac{}{}{0pt}{2}{\vec{k}_1+
    \vec{k}_2=\vec{k}_3+\vec{k}_4}{r,s,r',s'}}a^\dagger_{r'}(\vec{k}_1)
  a^\dagger_r(\vec{k}_2)V^{r's'}_{rs}a_s(\vec{k}_3)a_{s'}
  (\vec{k}_4),
\end{multline}
where $a_r\adj(\vec{k})$ and $a_r(\vec{k})$ are creation and
destruction operators, respectively, of one-particle plane wave states
with momentum $\vec{k}$ and spin projection $r$. The spin index $r$
refers to the eigenvalue of the \z-component of the spin operator and
can take the values $+,0,-$. Summation over repeated indices is
understood throughout the paper. In this usual basis the spin
operators are given by $F_z=\diag(1,0,-1)$ and $F_x=(F_++F_-)/2$ and
$F_y=(F_+-F_-)/2i$, and finally:
$F_\pm=\sqrt{2}(\delta_{r,\pm}\delta_{s,0}+\delta_{r,0}
\delta_{s,\mp})$. In Eq.  \eqref{eq:ham} $\kink=\hslash^2k^2/2M$
refers to the kinetic energy of an atom ($M$ is the mass of an atom),
$\mu$ denotes the chemical potential. The quantities $\tilde B$ and
$\olt$ are introduced as
\begin{equation}
  \label{eq:multiplikator}
  g\mb \tilde{B}\equiv \holt=\hol+g\mu_B\mu_m.
\end{equation}
Here $\hslash\ol$ is the Zeeman energy shift in a magnetic field: $\hol=g\mb
B$, where $g$ is the gyromagnetic ratio, $\mu_{\mathrm{B}}$ is the
Bohr magneton; $B$ is the modulus of the homogeneous magnetic
field and $\mu_m$ plays the role of a Lagrange multiplier for the
magnetization. In Eq.  \eqref{eq:ham} $V^{r's'}_{rs}$ is the Fourier
transform of the two particle interaction potential, which for the low
temperature, dilute gas can be modeled by the momentum independent
amplitude given for spin-1 bosons by \cite{Ho2,OM,Stamper-Kurn2001a}:
\begin{equation}
  \label{eq:pseudopot}
    V^{r's'}_{rs}=c_n\delta_{rs}\delta_{r's'}+c_s(\vec{F})_{rs}
    (\vec{F})_{r's'},
\end{equation}
In this paper we consider systems with $c_s>0$. An example of such a
system is the gas of \Na atoms \cite{Crea}. 

It is important that preparing the system with a suitable
magnetization the effective field $\tilde B$ can be made much smaller than the
external magnetic field $B$. This procedure made possible to observe
experimentally the phase P2 (see Fig. 3c in Ref.  \cite{Sea2} and
Figs. 21c, 22c in Ref. \cite{Stamper-Kurn2001a}). Note that the
quadratic Zeemann shift makes the phases P1 and P2 unstable for
increasing magnetic field. It is assumed throughout this paper that
the system is away from this stability border and the condensation
occurs only in the spin directions $+$ and $-$.

The Hamiltonian \eqref{eq:ham} is invariant under the gauge
transformations $a_{\pm}\rightarrow a_{\pm}e^{i\ph_{\pm}}$ and $a_0
\rightarrow a_0 e^{i\frac{1}{2}(\ph_++\ph_-)}$. This is equivalent to
$a_r\rightarrow a_re^{i(\phi+r\ph)}$, with $\phi=(\ph_++\ph_-)/2$ and
$\ph=(\ph_+-\ph_-)/2$. The invariance under transformations with
$\ph\equiv0$ yields a conservation law for the particle number, while
with $\phi\equiv0$ results in the conservation of the \z-component of
total magnetization.

We investigate such a parameter region $(\mu,\tilde B,T)$, where the system has a
Bose--Einstein condensate
$\avg{a_r(\vec{0})}=\sqrt{N_{\text{c}}}\zeta_r$, with $N_{\text{c}}$
the number of atoms in the condensate and
$\zeta_r=(\zeta_+,\zeta_0,\zeta_-)$ being the normalized spinor of the
condensate \cite{Ho2}. One can take  $\zeta_0=0$, which can be shown 
to be consistent with the general theory,
i.e. to all orders in the perturbation expansion. Note that when the
magnetic field and the magnetization are zero the magnitudes of the two
components of the spinor ($\zeta_+$ and $\zeta_-$) are equal. This
spinor is equivalent to what is usually taken for the polar state,
i.e. $\zeta=(0,1,0)$ \cite{Ho2}.  It is convenient to define a new set
of creation and annihilation operators: $b_r(\vec{k}) \equiv
a_r(\vec{k}) - \sqrt{N_{\text{c}}} \zeta_r \delta_{\vec{k},0}$ and
consider the Hamiltonian \eqref{eq:ham} as expressed with this new
set. The canonical transformation, together with the requirement
\begin{equation}
  \label{eq:canreq}
  \avg{b_r(\vec{k})}=\avg{b\adj_r(\vec{k})}=0
\end{equation}
let us to define an ensemble with density matrix $\rho = \exp(- \beta
\mathcal{H}) / Z$ that exhibits the symmetry breaking associated with
Bose--Einstein condensation.

\begin{figure}[t!]
  \centering
  \includegraphics[scale=0.83]{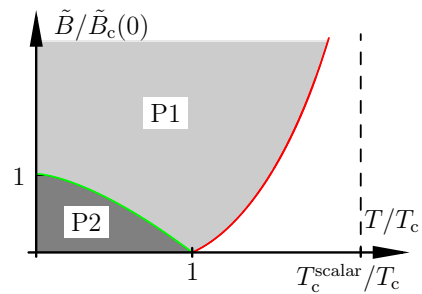}
  \caption{The phase diagram of the antiferromagnetic spin-1 Bose gas
    obtained in the mean field approximation}
  \label{fig:phdiag2d}
\end{figure}

The general features of the phase diagram can be read in Fig.
\ref{fig:phdiag2d} (calculated in an approximation to be introduced in
the second part of the paper). The red line designates the border for
BEC of the spin-1 gas, $\Tc$ denoting the critical temperature at
$\tilde B=0$. In phase P2 the condensate is characterized by a two
component spinor $\zeta = (\zeta_+, 0, \zeta_-)$,
$|\zeta_+|^2+|\zeta_-|^2=1$, while in P1 $\zeta=(1,0,0)$. For the
P2--P1 phase transition the order parameter is
$\sqrt{N_{\text{c}}}\zeta_-$. In the P2 phase both symmetries
expressing the conservation of the particle number and that of the
\z-component of the magnetization are broken by fixing $\ph_+$ and
$\ph_-$. In the P1 phase like in the ferromagnetic case \cite{Ho2} the
only remaining phase $\ph_+$ is related to the breaking of the combined
symmetry of particle number conservation and magnetization.

The key quantity in our presentation is the finite-temperature Green's
function of the system:
\begin{equation}
  \label{eq:grdef}
      \Gr^{rs}_{\gamma\delta}\ktau=-\Big<T_\tau\big[b^\gamma_r\ktau
    b^{\delta\adj}_s(\vec{k},0)\big]\Big>.
\end{equation}
$\tau$ is the imaginary time and $T_\tau$ refers to the $\tau$
ordering operator \cite{SG,Griffin}. The Greek indices take the values
 $\pm1$, with $b^\gamma_r(\vec{k})=b_r(\vec{k})$ if $\gamma=1$
and $b^\gamma_r(\vec{k})=b_r\adj(-\vec{k})$ if $\gamma=-1$.  Here
$\Gr^{rs}_{1,1}\ktau$ and $\Gr^{rs}_{-1,-1}\ktau$ stand for the normal
Green's functions, while $\Gr^{rs}_{1,-1}\ktau$ and
$\Gr^{rs}_{-1,1}\ktau$ for the anomalous ones, which arise due to
Bose--Einstein condensation.

The propagator $\Gr^{00}_{\gamma\delta}$ describes dynamics corresponding to
rotations around axes lying in the $x$-$y$ plane. Such  spin-wave
modes have a gap due to the nonzero effective magnetic field $\tilde B$.  One can
convince oneself, that the specific choice of the spinor, i.e.
$\zeta_0=0$, and of the direction of the magnetic field leads to $\Gr^{rs}_{\gamma\delta}=0$
for $r=0$ ($s=0$) and $s=\pm$ ($r=\pm)$. As a consequence $\Gr^{00}_{\gamma\delta}$
decouples from the other Green's functions and will not be discussed
here further since we concentrate on the order parameter related to
the P1--P2 transition and to its correlations described by
$\Gr^{--}_{\gamma\delta}$.The structure and couplings of the Green's function
$\Gr^{--}_{\gamma\delta}$ are different in the phases P1 and P2. The main results are
as follows.

In P2 $\Gr^{--}_{\gamma\delta}$ is coupled to $\Gr^{++}_{\gamma\delta}$ by $\Gr^{+-}_{\gamma\delta}$.
The $4\times4$ matrix with elements $\Gr^{rs}_{\gamma\delta}$, $r,s=\pm$ denoted by
$\mat{\Gr}$ takes the form in the Matsubara representation
\begin{equation}
  \label{eq:grpm}
  \mat{\Gr}^{-1}\komega=\Gro^{-1}\komega-\mat{\Sigma}\komega,
\end{equation}
with the self-energy given by the following hypermatrix notation:
\begin{equation}
  \label{eq:sepmd}
  \mat{\Sigma}=
  \left[
    \begin{array}{l r}
      \underline{\Sigma}^{++}& \underline{\Sigma}^{+-}\\
      \underline{\Sigma}^{-+}& \underline{\Sigma}^{--}
    \end{array}
  \right], \quad
  \underline{\Sigma}^{rs}=
  \left[
    \begin{array}{l r}
      \Sigma^{rs}_{1,1}& \Sigma^{rs}_{1,-1}\\
      \Sigma^{rs}_{-1,1}& \Sigma^{rs}_{-1,-1}
    \end{array}
  \right].
\end{equation}
The inverse of the free propagator is $\Gro^{-1} = \diag(\D+ \olt,
\WD + \olt, \D - \olt,\WD-\olt)$, with $\D = i\omega_n - \hslash^{-1}
(\kink - \mu)$. The asterisk denotes complex conjugation. 

Concerning the related spectrum it is important that the two
components of the condensate are independent in the sense that they
have independent phases: $\zeta' = (e^{i\ph_+} \zeta_+ , 0 ,
e^{i\ph_-}\zeta_-)$ is equivalent to $\zeta=(\zeta_+,0,\zeta_-)$.
Consequently two gapless Goldstone modes are expected. In this
respect it is important that generalized Hugenholtz--Pines (GHP)
theorems are valid. They are as follows
\begin{subequations}
  \label{eq:hupi}
  \begin{align}
    \Sigma^{++}_{1,1}(\vec{0},0)-\Sigma^{++}_{1,-1}(\vec{0},0)=
    \hslash^{-1}\mu+\olt,\label{eq:hupi1}\\
    \Sigma^{--}_{1,1}(\vec{0},0)-\Sigma^{--}_{1,-1}(\vec{0},0)=
    \hslash^{-1}\mu-\olt,\\
    \Sigma^{+-}_{1,1}(\vec{0},0)-\Sigma^{+-}_{1,-1}(\vec{0},0)=0,
  \end{align}
\end{subequations}
One can show that these relations provide necessary
conditions for the existence of gapless modes, since the denominator
of the Green's function matrix as defined in Eq. \eqref{eq:grpm} is
zero at $k=0$, $i\omega _n=0$ if Eqs. \eqref{eq:hupi} are fulfilled.
\begin{figure}[bt!]
  \centering
  \includegraphics[scale=0.83]{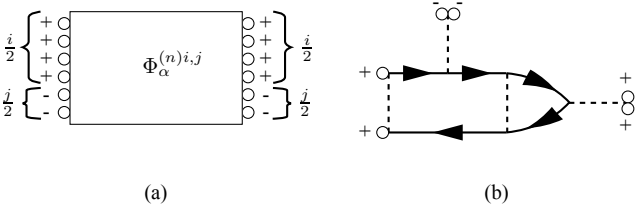}
  \caption{(a) The symbolic representation of an irreducible diagram
    having $i$ condensate lines with $r=+$ and $j$ with $r=-$. (b) A
    fourth order example is shown with $i=4$ and $j=2$. The dashed line represents
   the interaction, the circle stands for the square root of the corresponding condensate density
   and the line with an arrow denotes the free Green's function.}
  \label{fig:vacdiag}
\end{figure}

We sketch the derivation, because it sheds light on the structure of
the perturbation expansion. Namely, all $n^{\text{th}}$ order diagrams
contributing to  the self-energies $\Sigma^{rs}_{11} (\vec{0}, 0)$ and $\Sigma^{rs}_{1,-1} (\vec{0}, 0)$
can be obtained easily from the one-particle irreducible graphs
without external lines but with $i$ number of condensate lines with
spin projection $+$ and $j$ number of condensate lines with spin
projection $-$. Particle number and spin
conservation results in $i$ and $j$ being even. For an illustration of
such a graph see Fig.  \ref{fig:vacdiag}, where the condensate lines
are represented by circles. Let us denote its
contribution by $\Phi^{(n)i,j}_{\alpha}$. Here $n$ refers to the order
in perturbation theory and the index $\alpha$ is for distinguishing
between graphs with the same $n,i,j$ but different topologies and
consequently different contributions. To obtain the $n^{\text{th}}$
order contribution of the normal self-energy $\Sigma^{++}_{11} (\vec{0}, 0)$ one has to replace two
appropriately chosen condensate lines by an incoming
and an outgoing particle line.  Note that those circles where incoming lines can be
substituted are gathered at the left side of Fig. \ref{fig:vacdiag} (a) while those 
standing for possible places of outgoing lines at the right side.
Finally one arrives at
\begin{equation}
  \label{eq:sigma++11}
  \Sigma^{++}_{11}(\vec{0},0)=\frac{1}{\ncp}\sum_i\sum_\alpha\bigg(\frac{i}{2}
  \bigg)^2\Phi^{(n)i,j}_{\alpha},
\end{equation}
where $n_{\text{c},r}=N_\text{c}|\zeta_r|^2/V$ is the condensate
density in spin projection $r\in\{+,-\}$. For obtaining the anomalous $\Sigma^{++}_{1-1} (\vec{0}, 0)$
self-energy one has to replace two appropriately chosen condensate lines  by two outgoing lines. One gets that
\begin{equation}
  \label{eq:sigma++1-1}
  \Sigma^{++}_{1,-1}(\vec{0},0)=\frac{1}{\ncp}\sum_i\sum_\alpha
  \bigg(\frac{i}{2}\bigg)\bigg(\frac{i}{2}-1\bigg)\Phi^{(n)i,j}_{\alpha}.
\end{equation}

It is possible to calculate this way the contribution of the
irreducible tadpole diagrams as well, i.e. those with one incoming
particle line:
\begin{equation}
  \label{eq:sigma+0}
  \Sigma^{+}_{0}=\frac{1}{\sqrt{\ncp}}\sum_i\sum_\alpha
  \bigg(\frac{i}{2}\bigg)\Phi^{(n)i,j}_{\alpha}.
\end{equation}
Using Eqs. \eqref{eq:sigma++11}, \eqref{eq:sigma++1-1} and
\eqref{eq:sigma+0} one arrives at
\begin{equation}
  \label{eq:hupiproof1}
  \Sigma^{++}_{11}(\vec{0},0)-\Sigma^{++}_{1,-1}(\vec{0},0) = 
\frac{1}{\sqrt{\ncp}}\Sigma^{+}_{0}.
\end{equation}
The requirement formulated in Eq. \eqref{eq:canreq} for $r=+$ is satisfied when
$\Sigma^{+}_{0}=\hslash^{-1}\mu+\olt$, that yields with
\eqref{eq:hupiproof1} one of the GHP theorems \eqref{eq:hupi1}. One
can prove the others similarly.

Crossing the border line one enters the P1 phase where the condensate
spinor has only one nonzero component $\zeta=(1,0,0)$. This spinor has
only one phase, which can be chosen freely; a continuous symmetry is restored and
therefore only one Goldstone mode exists. Furthermore in phase P1
$\Sigma^{+-}_{\alpha\gamma} = \Sigma^{-+}_{\alpha\gamma}=0$.
Consequently $\Gr^{++}_{\alpha\gamma}$ and $\Gr^{--}_{\alpha\gamma}$ are no more coupled. The
remaining Goldstone mode shows up in the spectrum of $\Gr^{++}_{\alpha\gamma}$ and
the HP theorem $\Sigma^{++}_{1,1}(\vec{0},0) - \Sigma^{++}_{1,-1}
(\vec{0},0) = \hslash^{-1} \mu + \olt$ ensures its gapless nature.  The
order parameter dynamics as described by $\Gr^{--}_{\alpha\gamma}$ further simplifies
into 
\begin{equation}
  [\Gr^{--}_{11}\komega]^{-1}=i\omega_n-\hslash^{-1}(\kink-\mu+
    \holt)-\Sigma^{--}_{11},
  \label{eq:propfer-}
\end{equation}
since $\Sigma^{--}_{1,-1}=0$ \cite{KSzSz2}.
The excitation spectrum determined by Eq.  \eqref{eq:propfer-} has a
gap, which disappears when reaching the phase boundary between P1 and
P2:
\begin{equation}
  \label{eq:crmf}
  \holt^{\mathrm{c}}=\mu-\hslash\Sigma^{--}_{11}(\vec{0},0)\big|_{\olt=
    \olt^{\mathrm{c}}}.
\end{equation}

At $T=0$ this condition designates a quantum phase transition and it
is expected that there is a crossover from classical to quantum phase
transition when $T \rightarrow 0$ \cite{Sondhi1997a}. One can define
the shift exponent $\ps$ by the equation $\tilde\BC(T) = \tilde\BC(0)
- w T^{1/\ps}$ \cite{Fisher1974a}. It is useful then to introduce the
variable $x=T/[\tilde B-\tilde\BC(0)]^{\ps}$. Denoting by $\xc$ its
value on the critical line, one can write the leading singularities as
follows
\begin{align}
  \label{eq:g--scale}
  \Gr^{--}_{11}(\vec{0},0)&=\frac{C(T)}{\left|x-\xc\right|^{\gamma}},
  \quad T>0,\\
  &=\frac{Q}{\left|\tilde B-\tilde\BC(0)\right|^{\gamma_0}},
  \quad T=0.\nonumber
\end{align}
One can show that $\ps$ agrees with the usually defined crossover
exponent $\phi$
when crossover scaling is valid [in this case $C(T)\sim
T^{-\gamma_0/\phi}$ in Eq.  \eqref{eq:g--scale}].
One can specify a low temperature region where the energy gap is
higher than the temperature.
Here only exponentially small corrections  due to nonzero
temperatures are expected.

In the following we will discuss static and dynamic properties within
a kind of a mean field approach. We start by writing self-consistent
equations for the densities of the conserved quantities
\begin{subequations}
  \label{eqs:heqs}
  \begin{align}
    n&=\ncp+\ncm+n'_++n'_0+n'_-,\\
    m&=\ncp-\ncm+n'_+-n'_-,\label{eq:heq2}
  \end{align}
\end{subequations}
with $n$ being the total particle density and $m$ the magnetization
density. The density of non-condensed atoms in spin projection
$s\in\{+,0,-\}$ is given by the Bose distribution
$n'_s=V^{-1}\sum_{\vec{k}}n'_{\vec{k},s}$, with
$n'_{\vec{k},s}=(e^{\beta\eH_{\vec{k},s}}-1)^{-1}$ and
\begin{equation}
  \label{eq:eharpa}
  \eH_{\vec{k},s}=\kink+c_n n-\mu+s(c_s m - \holt).
\end{equation}
There exists also a relation following from the requirement
$\avg{b_+(\vec{0})}=0$, which provides the equation
\begin{subequations}
  \begin{equation}
    \label{eq:chempot1}
    \mu+\holt=c_n n + c_s m.
  \end{equation}
  This equation holds both in the P1 and P2 phases. In phase P2
  $n_{0,-}\neq0$ and a second equation emerges from the requirement
  $\avg{b_-(\vec{0})}=0$, which reads as
  \begin{equation}
    \label{eq:chempot2}
    \mu-\holt=c_n n - c_s m.
  \end{equation}
\end{subequations}
Using Eqs.  \eqref{eqs:heqs}, \eqref{eq:eharpa}, \eqref{eq:chempot1}
and \eqref{eq:chempot2} one can get the form of the spinor in phase
P2, which reads
\begin{equation}
  \label{eq:spinor}
  \zeta_{\pm}=\left[\frac{1}{2}\left(1\pm\frac{\holt}{c_sn_c}\right)\right]^\frac{1}{2},
\end{equation}
where $n_c$ is the total density of the condensate.

The obtained phase diagram is depicted in Fig. \ref{fig:phdiag2d}.
For the critical line one gets
\begin{equation}
  \label{eq:critline}
  \frac{\tilde B_{\text{c}}(T)}{\tilde B_{\text{c}}(0)}=
  1-\left(\frac{T}{T_c}\right)^{3/2},
\end{equation}
with
$T_c=(4\pi^2n)^{2/3}\hslash^2/(3\Gamma(3/2)\zeta(3/2))^{2/3}2Mk_B$
being the critical temperature at $\tilde B=0$. Therefore the shift
exponent is $\ps=2/3$.

The second step is to express the Green's functions in harmony with
the equation of state as written above. It can be shown that the
self-consistent nature of the equation of state and the existence of
the condensate leads to RPA like contributions to the self-energies.
To see that the procedure leads to a conserving approximation one
needs to treat correlation functions of the density and spin density
operators in detail, which goes beyond the scope of the present paper
and will be published together with the dielectric formalism
generalized to $\tilde B\neq0$ \cite{KSzSz2}.

We consider the P1 phase first. Since
$\ncm=0$ in this phase the RPA like terms are not present in
$\Sigma^{--}_{\gamma\delta}$ (we recall that $\Sigma^{+-}_{\gamma\delta}=0$ anyhow in P1).
Consequently $\Gr^{--}_{11} = [i\omega_n - \hslash^{-1}
\eH_{\vec{k},-}]^{-1}$ which after a somewhat lengthy calculation
leads to \eqref{eq:g--scale} with $\gamma=2$ and $\gamma_0=1$ and
furthermore to $C(T)\sim T^{-1}$. Further critical exponents have the values
$\eta=\eta_0=0$, $\nu=1$, $\nu_0=1/2$. They are defined as follows:
$\Gr^{--}_{11}(\vec{0},0)\sim1/k^{2-\eta}$ on the critical line, while
$\Gr^{--}_{11}(\vec{0},0)\sim1/k^{2-\eta_0}$ at the quantum critical point;
the correlation length $\xi\sim1/|x-x_c|^\nu$ and $\xi\sim1/|\tilde B-\tilde\BC(0)|^{\nu _0}$
at $T>0$ and at $T=0$, respectively.
The dispersion relation is real and parabolic with a gap, that
vanishes at the phase boundary between P1 and P2. The gap exponent is
$\nu z$ ($T>0$), $\nu_0 z$ (T=0), where the dynamical scaling exponent
$z=2$. Choosing a path towards the quantum critical point such that
$[\tilde B-\tilde\BC(0)]/T = \text{const} > 0$, one obtains
$\Gr^{--}_{11}(\vec{0},0) \sim T^{-\nu_0 z}$. The slope of the border line
of the low temperature region [specified below Eq.
\eqref{eq:g--scale}] is proportional to $1/c_s$ and is about 200 in
the units of Fig.  \ref{fig:phdiag2d} by choosing  the following realistic parameter
values: $c_n=1.03 \cdot 10^{-50} \mathrm{J\,m^3}, c_s=3.21 \cdot
10^{-52} \mathrm{J\,m^3}$ and the density
($4.6\cdot10^{20}\mathrm{m}^{-3})$.  The correction to the leading
$1/T$ singularity has a weaker power law dependence on $1/T$ (in the model $T^{-1/2}$).
This region, which is presumably experimentally accessible terminates
at the line $\tilde B=\tilde \BC(0)$ along with
$\Gr^{--}_{11}(\vec{0},0)\sim T^{-3/2}$.  Note that while the power law
behaviours are expected to be generally valid the exponents along the
critical line are characteristic for the model. Namely, they are those
of the spherical model. The reason is that the model can be formally  derived by
introducing $\sigma$ species of spin-1 bosons and taking the limit
$\sigma\rightarrow\infty$ while the interaction parameters $c_n$,
$c_s$ are going to zero as $1/\sigma$.  Furthermore $B$, $\mu$ and
$\mu _m$ remain of ${\mathcal O}(1)$.

In phase P2 the complete self-energy matrix \eqref{eq:sepmd} has to be
treated.  One obtains for the Green's function \cite{KSzSz2}
\begin{multline}
  \left(\mat{\Gr}^{-1}\right)^{rs}_{\alpha\gamma}=\left(i\omega_n-
    \hslash^{-1}\kink\right)\delta_{rs}\delta_{\alpha\gamma}\\
  -\hslash^{-1}\sqrt{n_{\text{c},r}n_{\text{c},s}}
   (\cn+rs\cs),
 \end{multline}
 for $r,s\in\{+,-\}$. The effective interactions are given by
 $\cn\komega = c_n/[1-3c_n\Pio\komega]$,
 $\cs\komega=c_s/[1-2c_s\Pio\komega]$.  The polarization function
 $\Pio$ is the contribution of the bubble diagram of the free gas,
 since $\eH_{\vec{k},s}=\kink$ in P2 as can be seen from Eqs.
 \eqref{eq:eharpa}, \eqref{eq:chempot1} and \eqref{eq:chempot2} and
 reads as 
 \begin{equation}
   \label{eq:bubble}
   \Pio\komega=-\frac{1}{\hslash}\int\frac{{\mathrm d^3}q}{(2\pi)^3}
   \frac{n'_{\vec{k}+\vec{q},s}-n'_{\vec{q},s}}
   {i\omega_n-\hslash^{-1}(e_{\vec{k}+\vec{q}}-e_{\vec{q}})},
 \end{equation}
for all values of $s$.
By writing $c_n,c_s$ instead of $\cn,\cs$ the Bogoliubov approximation 
of the Green's functions is formally recovered. It is
remarkable that the GHP theorems \eqref{eq:hupi} are fulfilled in
this model in such a way that simultaneously
\begin{equation}
  \label{eq:nngen}
  \Sigma^{rs}_{\alpha\beta}(\vec{k},0)=\Sigma^{rr}_{\alpha\alpha}(\vec{k},0)
  \delta_{rs}\delta_{\alpha\beta},\qquad k\rightarrow0.
\end{equation}
Repeated indices are not summed in this equation. If \eqref{eq:nngen}
were true in general then it would mean a generalization of the
results by Nepomnyashchi{\u\i} and Nepomnyashchi{\u\i} \cite{NN1} derived at
$T=0$ in the case of liquid Helium, i.e. for particles with zero spin.
Namely, by analyzing the infrared divergences they have
shown that the anomalous self-energy $\Sigma_{1,-1}(\vec{0},0)=0$. At
nonzero temperature the expected behaviour is that
$\Sigma_{1,-1}(\vec{k},0)\sim k$ \cite{SG}, which is valid here for
all the nondiagonal self-energies, since $\Pio(\vec{k},0)\sim k^{-1}$.

\begin{figure}[t!]
  \centering
  \includegraphics[scale=0.5]{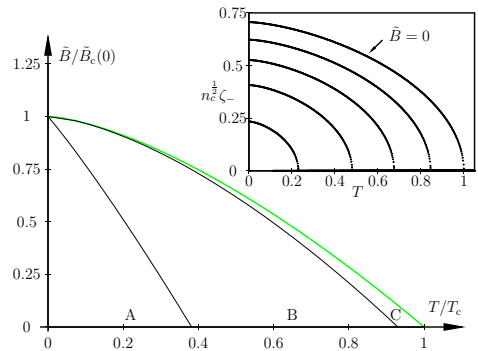}
  \caption{The three regions of the phase P2. Region A and C are
    enlarged for the sake of better visualization. The inset shows the
    order parameter as a function of temperature for several values of
    $\tilde B$.}
  \label{fig:region}
\end{figure}

The denominator of the Green's function (after analytical continuation
in frequency) reads as:
\begin{multline}
  \label{eq:grdenp2}
  \Delta(\vec{k},\omega)=\big[\omega^2-\hslash^{-2}\kink^2\big]^2(1-
  2c_s\Pior)(1-3c_n\Pior)\\
  -2\hslash^{-2}\kink\big[\omega^2-\hslash^{-2}\kink^2\big]\nc
  (c_n+c_s-5c_nc_s\Pior)\\
  +16\hslash^{-4}\kink^2c_nc_s\ncp\ncm.
\end{multline}
The spectrum of elementary excitations is given by
$\Delta(\vec{k},\omega) |_{\omega=E_{\vec{k}} - i\gamma_{\vec{k}}}=
0$. Three correlation lengths will be used to specify regions in the
phase diagram with qualitatively different solutions, namely the
thermal wavelength $\lambda=\hslash/(2M\kb T)^{1/2}$, the Bogoliubov
coherence length $\xiB=\hslash/(4Mc_s\ncm)^{1/2}$, and the length
characterizing phase fluctuations of the order parameter $\xi'=M\kb
T/(4\pi\hslash^2\ncm)$. The three regions of the phase P2 are
illustrated in Fig \ref{fig:region}. For the sake of simplicity we restrict the
discussion to small $k$ values and assume that $c_s/c_n$ is small,
that is the case in the experiment \cite{Stamper-Kurn2001a}.
\emph{Region A} ($\lambda\gg\xiB,\xi'$): The polarization function can
be approximated as
\begin{equation}
\Pior(\vec{k},\omega)=\frac{2p\kink}{\omega^2},
\end{equation}
where $p=\Gamma(3/2)\zeta(3/2)/4\pi^2\lambda^3$.  The excitation
energies are linear in wavenumber: $\omega_{\pm}=c_{\pm}k$. Here
$c_+^2 = n c_n/M - 4c_sn_+n_-/(n M)$ and $c_-^2 = n_0
c_s/M+4c_sn_+n_-/(n M)$, where $n_r=n_{\text{c},r}+n_r'$.  At $T=0$
the result reduces to that of Ohmi and Machida obtained using the
equation of motion method \cite{OM}.  In the present approximation a
damping appears only for nonzero temperature, and is exponentially
small. In higher order a Beliaev type damping can arise (see \cite{SG}
for a detailed investigation of such a damping in case of a gas of
spinless bosons).  It is assumed that $\tilde B$ is large enough that the
interesting region in P2 keep away from the critical line of BEC.
Then $\omega_+$ does not exhibit any remarkable change when
approaching the P2-P1 boundary and therefore only $\omega_-$ will be followed
below.  In \emph{region B} ($\xiB \gg \xi', \lambda$) the polarization
function is approximated as
\begin{equation}
\Pior(\vec{k},\omega)
=-i\frac{\hslash}{2Mc_s}\frac{\xi'}{\xi_B^2}\frac{k}{\omega}.
\end{equation}
The velocity contains only the condensate densities
\begin{subequations}
  \begin{equation}
    c_{-}^2=\frac{4\ncp\ncm}{\nc}\frac{c_s}{M}
  \end{equation}
  and the mode has a Landau type damping
  \begin{equation}
    \label{eq:regionBdamp}
    \gamma_\vec{k}=\frac{Mc_s}{4\pi\hslash^3}\left[2+3
      \left(\frac{\holt}{c_s\nc}\right)^2\right]\kb T k.
  \end{equation}
\end{subequations}

\emph{Region C} ($\xi' \gg \xiB,\lambda)$ is the critical region. Here
the polarization function $\Pior\komega$ takes the same form as in
region B. For $\tilde B,T$ fixed when $k\rightarrow0$ the mode becomes
overdamped and soft, $\omega_-=-i2\hslash k / (5M\xi')$.  For $k$ fixed
$\omega_-$ tends to $\kink$ when the critical line is approached.  The
size of these regions depends crucially on $c_s$ for the other
parameters fixed. For realistic values of parameters the region A and C are very
narrow, the region B is enlarged and experimentally accessible. Here
the velocity depends on the condensate densities; the thermal
excitations represent higher order corrections. More
characteristically the damping is proportional to the wave number and
the temperature. These features are known for density waves in Bose
gases with frozen internal degrees of freedom \cite{SzK,SG,FRSzG} and
have been obtained also for spinor Bose gas at zero magnetization for
the density (and spin density) fluctuations \cite{SzSz2}. The linear
dependence on $T$ is even valid for gases in a magnetic trap
\cite{PS,FSW,Rea1}.  New feature of \eqref{eq:regionBdamp} is the
dependence on the strength of the magnetic field and magnetization.
One expects that the leading term remains quadratic in $\tilde B$
also for a gas in an optical trap.

In conclusion, we have investigated  the equilibrium and
dynamic properties of the two Bose--Einstein condensed phases of a spin-1
Bose gas.  The richness of properties of the two phases from the
point of view of a many body problem is demonstrated by the GHP theorems.
They reflect the basic fact that the number of Goldstone modes agrees with the
independent phase fluctuations in the complex plane exhibited by the
order parameter.  One of them (in P2) is the critical mode for the
P2--P1 transition, the other will become critical when the temperature
in P1 reaches the line of BEC. The first mode (which coincides with
the transverse spin wave only when the magnetization is zero) becomes
a quadrupolar spin wave and develops a gap in P1. It has been shown
that the transition between the phases P1 and P2 at zero temperature
belongs to the category of quantum phase transitions in which a
critical line starts from quantum critical point when the temperature
is raised. The measurement of the shift exponent would be of
particular interest.

The work was supported by the Hungarian National Research Foundation
under Grant No. OTKA T046129.

\end{document}